\documentclass{article}

\usepackage[utf8]{inputenc}
\usepackage{amsmath, amsthm, amssymb, amsfonts, dsfont, mathrsfs, dirtytalk, tikz-cd, adjustbox, url, nicefrac, booktabs,array, graphicx,todonotes,longtable, tabu,bm,cite, todonotes,comment,multirow,tabularx,epsfig,parskip}
\usepackage[T1]{fontenc}    
\usepackage[margin=1in]{geometry}
\usepackage[english]{babel}
\usepackage{hyperref}
\usepackage[capitalise]{cleveref}

\newcommand{\f}{\forall}

\theoremstyle{plain}

\theoremstyle{definition}
\newtheorem{definition}{Definition}

\theoremstyle{remark}

\title{Towards a Bayesian mechanics of metacognitive particles: A commentary on “Path integrals, particular kinds, and strange things” by Friston, Da Costa, Sakthivadivel, Heins, Pavliotis, Ramstead, and Parr}
\author{Lancelot Da Costa$^{1,2,3}$, Lars Sandved-Smith$^{4,}\thanks{Correspondence: \url{lars.sandvedsmith@gmail.com}}$ \\\\
$^1$\textit{Department of Mathematics, Imperial College London} \\
$^2$\textit{Wellcome Centre for Human Neuroimaging, University College London}\\
$^3$\textit{VERSES AI Research Lab}\\
$^4$\textit{Monash Centre for Consciousness and Contemplative Studies}}
\date{\vspace{-10pt}}

\begin{document}

\maketitle

Following the typology of particles introduced in the target article \cite{fristonPathIntegralsParticular2023}, we introduce a further distinction between cognitive and metacognitive particles. Metacognitive particles are called as such because they 
have beliefs about (some of) their own beliefs.

\begin{figure}[h]
\centering\includegraphics[width=0.7\textwidth]{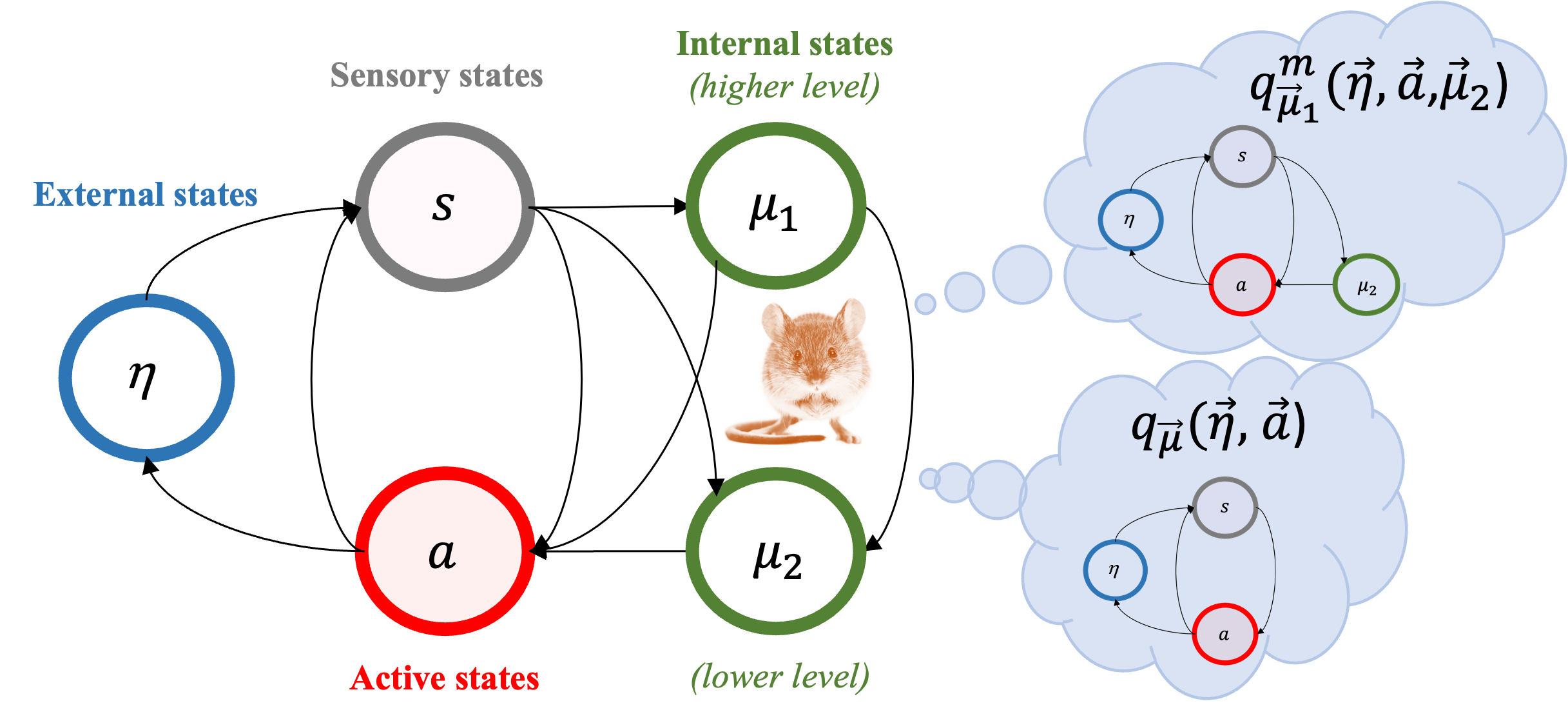}
\caption{\textbf{Example of a metacognitive strange particle.} This particle is such that internal states (at the lower level, i.e. $\mu_2$) are hidden from internal states (at the higher level, i.e. $\mu_1$). In other words, higher level internal states must infer the lower level internal states via the sensory states, leading to the metacognitive belief $q_{\vec{{\mu}}_1}^m$.}
\label{fig: metacognitive particle}
\end{figure}

The target article argues that strange particles may be most apt for describing sentient behaviour \cite{fristonPathIntegralsParticular2023}.
So let's consider a strange particle whose internal state-space can be decomposed into two sets of states $\mu\triangleq(\mu_1,\mu_2)$, such that the second set of states is hidden from the first via the Markov blanket. That is, $\mu_1$ may influence $\mu_2$ directly, but $\mu_2$ may only influence $\mu_1$ via the Markov blanket. To visualize this, this is the flow governing a metacognitive particle:
\begin{equation}
\label{eq: flow}
    \begin{split}
        f(\eta, s, a,\mu)=\begin{bmatrix}f_\eta(\eta, s, a) \\ f_s(\eta, s, a) \\ f_a(s, a, \mu) \\ f_{\mu_1}(s, \mu_1)\\f_{\mu_2}(s, \mu)\end{bmatrix}
    \end{split}
\end{equation}
The form of the coupling is summarised in Figure \ref{fig: metacognitive particle}. 
We will call $\mu_1$ the \textit{higher}-level internal states and $\mu_2$ the \textit{lower}-level internal states in virtue of the fact that $\mu_1$ will be seen as encoding beliefs at a higher level, i.e. some sort of metacognition.

Analogously to the discussion on strange particles \cite[eq. 29]{fristonPathIntegralsParticular2023}, the form of the coupling \eqref{eq: flow}, means that sensory states are a Markov blanket between higher- and lower-level internal states:
\begin{equation}
(\vec{\mu}_1 \perp \vec{\eta}, \vec{a}, \vec{\mu}_2) \mid \vec{s}
\end{equation}
In particular, the lower internal states can only be inferred vicariously by the higher internal states via the sensory states. So we can define \textit{metacognitive} beliefs encoded by the higher-level internal states $q_{\vec{{\mu}}_1}^m(\vec{\eta}, \vec{a},\vec{\mu}_2)$, where\footnote{These are well defined for a large class of strange particles in virtue of conditions analogous to that discussed in \cite[p. 24]{fristonPathIntegralsParticular2023}.} 
\begin{equation}
\label{eq: meta}
\begin{split}
q_{\vec{\boldsymbol{\mu}}_1}^m(\vec{\eta}, \vec{a},\vec{\mu}_2)  &\triangleq p(\vec{\eta}, \vec{a} ,\vec{{\mu}}_2\mid \vec{s}) \\
\vec{\boldsymbol{\mu}}_1  &\triangleq \arg \max _{\vec{\mu}_1} p(\vec{\mu}_1 \mid \vec{s})
\end{split}
\end{equation}
This sort of beliefs encoded by the higher internal state about the lower internal state is interesting because it implies that the higher internal states encode beliefs about beliefs, licensing the metacognitive terminology. Indeed, recall that a particle holds beliefs $q_{\vec{{\mu}}}(\vec{\eta})$ about the external world encoded by the internal states, where by definition \cite{fristonPathIntegralsParticular2023},
\begin{equation}
\begin{split}
    q_{\vec{\boldsymbol{\mu}}}(\vec{\eta}) & = p(\vec{\eta} \mid \vec{s}, \vec{a}) \\
\vec{\boldsymbol{\mu}} & \triangleq \arg \max _{\vec{\mu}} p(\vec{\mu} \mid \vec{s}, \vec{a})
\end{split}
\end{equation}
(See \cite{dacostaBayesianMechanicsStationary2021a} for an explicit construction).
It then follows that metacognitive beliefs encode information about the parameters of beliefs about the world, evincing a form of \textit{parametric depth} \cite{sandved-smithComputationalPhenomenologyMental2021}. Perhaps the easiest way to see this is when the particular beliefs factorise according to a mean-field approximation:
\begin{equation}
    q_{\vec{\mu}}(\vec{\eta})=q_{\vec{\mu}_1}(\vec{\eta})q_{\vec{{\mu}}_2}(\vec{\eta})
\end{equation}
Then the (marginal) metacognitive belief $q_{\vec{{\mu}}_1}^m(\vec{\mu}_2)$ is a belief about the belief $q_{\vec{{\mu}}_2}(\vec{\eta})$. The upshot is that once we have beliefs about some of our own internal states, we are bound to exhibit a form of metacognition that influences behaviour. For example, a person who is surprised by their sudden inability to taste coffee (a possible sign of pancreatic cancer) might introspect and act accordingly \cite{hohwyThePredictiveMind2013}. In summary:

\begin{definition}[Cognitive and metacognitive particles]
    A \emph{metacognitive} particle is a particle which holds metacognitive beliefs, i.e. beliefs about beliefs, e.g. as in \eqref{eq: meta}. Any non-metacognitive particle, on the other hand, is a \emph{cognitive} particle, given that it holds beliefs about the world.
\end{definition}

Note also that we could study higher levels of metacognition, by considering particles with a unidirected chain of internal states $\mu_1\rightarrow \mu_2 \rightarrow \hdots \rightarrow \mu_n$, cf. Figure \ref{fig: metacognitive particle}. In this case the internal states at the highest level, i.e. $\mu_1$, may encode beliefs about beliefs about beliefs etc ($n$ times).

In conclusion, we believe that the typology of particles laid down in the target article \cite{fristonPathIntegralsParticular2023} is promising, not least for a physics of cognition that builds upon and refines the free energy principle \cite{fristonFreeEnergyPrinciple2023a}. As a result, we are starting to be able to derive formal descriptions of cognition that are increasingly specific to entities possessing higher forms of cognition, such as, we hope, the human being.

\section*{Funding statement}

L.D. is supported by the Fonds National de la Recherche, Luxembourg (Project code: 13568875). This publication is based on work partially supported by the EPSRC Centre for Doctoral Training in Mathematics of Random Systems: Analysis, Modelling and Simulation, United Kingdom (EP/S023925/1). L.S.S. is supported by the Three Springs Foundation.

\bibliographystyle{unsrt}
\bibliography{bib}

\end{document}